\documentclass[prb,preprint]{revtex4}
\usepackage{amsmath}

\usepackage{graphicx}

\begin{document}

\title{Walther Bothe and Bruno Rossi: The birth and development of coincidence methods in cosmic-ray physics}

\author{Luisa Bonolis}
\email{luisa.bonolis@roma1.infn.it}
\affiliation{Italian Association for the Teaching of Physics (A.I.F.)  History of Physics Group [Via Cavalese 13, 00135 Rome, Italy]}


\begin{abstract}
 In 1924 Walther Bothe and Hans Geiger applied a coincidence method to the study of Compton scattering with Geiger needle counters. Their experiment confirmed the existence of radiation quanta and established the validity of conservation principles in elementary processes. At the end of the 1920s, Bothe and Werner Kolh\"orster coupled the coincidence technique with the new Geiger-M\"uller counter to study cosmic rays, marking the start of cosmic-ray research as a branch of physics. The coincidence method
was further refined by Bruno Rossi, who developed a vacuum-tube device capable of registering the simultaneous occurrence of electrical pulses from any number of counters with a tenfold improvement in time resolution. The electronic coincidence circuit bearing Rossi's name was instrumental in his research on the corpuscular nature and the properties of cosmic radiation during the early 1930s, a period characterized by a lively debate between Millikan and followers of the corpuscular interpretation. The Rossi coincidence circuit was also at the core of the counter-controlled cloud chamber developed by Patrick Blackett and Giuseppe Occhialini, and became one of the important ingredients of particle and nuclear physics. During the late 1930s and 1940s, coincidences, anti-coincidences and delayed coincidences played a crucial role in a series of experiments on the decay of the muon, which inaugurated the current era of particle physics. 

\end{abstract}

\maketitle

\section{Introduction}
With the advent of Geiger-M\"uller  (G-M) counter) in the late 1920s, cosmic ray research changed dramatically: for the first time, the physical nature of cosmic rays became accessible to experimentation. However, when used single, as cosmic-ray detectors, these devices did not have significant advantages over ionization chambers. They became a most powerful new tool for cosmic-ray experiments when used in coincidence arrangements.  The coincidence technique, first used by Hans Geiger and Walther Bothe in 1924 to verify that Compton scattering produces a recoil electron simultaneously with the scattered $\gamma$-ray,  achieved its full potentialities only in connection with the invention of electronic circuits at the beginning of 1930s. From then on, in conjunction with the invention of new sophisticated  detectors, the coincidence method became one of the basic tools in the art of experimental physics. 

The following historical reconstruction, a scientific saga extending from the 1920s to late 1940s, will examine the scientific literature of the time in order to outline how the arrangement of complex arrays of counters, absorbers and electronic recording circuits became  standard  in cosmic-ray studies, as well as in nuclear and particle physics.

\section{Waves and corpuscles in the 1920s}

At the beginning of the 20th century the classical, continuous-wave picture of radiation was challenged by Planck's elementary quantum of action unexpectedly  generated by his theory of black-body spectrum, and especially by Einstein's light quantum interpretation of the photoelectric effect of 1905, whose validity was experimentally proved for the first time in 1916 by Robert Millikan.\cite{MillikanPais} The question was taken over again  in 1922 by Arthur Compton.  Between 1916 and 1922, Compton pursued an experimental and theoretical research that culminated in the discovery that, contrary to what he had expected, the wavelength of X-rays increased due to scattering of the incident radiation by free electrons. The X-rays behaved as particles capable of exchanging their energy and momentum with another particle (the electron) during collisions.
Peter Debye emphasized the importance of this discovery in support of light-quanta propagation.\cite{Compton}

Most physicists at that time believed that light quanta did not represent physical reality, and was just a heuristic way of defining some quantity of energy related to a property of electromagnetic fields.
Bohr expressed his opposition to the concept during his Nobel Lecture of 1922: ``In spite of its heuristic value the hypothesis of light quanta, which is quite irreconcilable with the so-called interference phenomena, is not able to throw light on the nature of radiation."\cite{Bohr1922}

Notwithstanding the close agreement between theory and the actual wavelengths of the scattered rays observed by Compton, there was no direct evidence for the existence of the recoil electrons required by the theory of light-quantum scattering. Within a few months of Compton's first official announcement, the cloud chamber gave strong support to the validity of the Compton process and to the particle interpretation of electromagnetic radiation.

In 1923 Charles Wilson perfected his device,\cite{Wilson1923} which Ernest Rutherford considered to be ``the most original and wonderful instrument in scientific history,''\cite{Rutherford} and took photographs of the tracks of fast electrons ejected from atoms by X-rays. After claiming that ``If each $\beta$-ray which is produced by the action of the X-rays represents the absorption of one quantum of radiation, the method enables us to deal directly with individual quanta,'' Wilson explicitly mentioned Compton's work.\cite{Wilson1923a}

Wilson's methods were quickly followed by success in many parts of the world,
in particular in Berlin, by Walther Bothe and Lise Meitner. Having been Max Planck's student, Bothe had a strong background in quantum theory, and in 1923 he was doing research on the corpuscular theory of light, as well as on the nature of radiation and the properties and behavior of electrons. In the months preceding the announcement of the Compton effect, he observed the short tracks of the Compton recoil electrons of X-rays in a cloud chamber filled with hydrogen.\cite{Bothe}

According to Compton's experiments, it appeared that energy was conserved during the collision process. However, it was not quantitatively determined. Compton observed only the X-rays after the collision with an electron, and determined that their energy had the values according to calculations which considered X-rays as particles, with their own energy and momentum. Compton did not detect the electron, and he did not know if the electron had gained all the energy lost by a single X-ray, or a part which  varied during the single scattering process.  In 1924 Bothe came across a theoretical paper by Bohr, Hendrik Kramers, and John Slater in which the authors tried to reconcile quantum effects with Maxwell's theory of the electromagnetic field.\cite{BKS1924} They had to accept that energy and momentum are not conserved in each individual process, but only in the statistical average.
Bohr was one of the opponents to the light-quantum nature of radiation. According to this hypothesis, there was one scattered light quantum for any recoiling electron and vice versa, whereas in the Bohr, Kramers, Slater theory, recoil would occur in any direction with nonzero probability. As Bothe recalled in his 1954 Nobel Lecture: ``In the individual or elementary process, so long as only a single act of emission was involved, the laws of conservation were held to be statistically satisfied only, to become valid for a macroscopic totality of a very large number of elementary processes only, so that there was no conflict with the available empirical evidence.''\cite{Bothe1954}

Bothe had been working in Berlin since 1913 at the Physikalisch-Technische Reichsanstalt, where Hans Geiger had just become its director.  Geiger was a pioneer in the counting technique  since his post-doctoral sojourn in Rutherford's laboratory between 1906 and 1912.\cite{Geiger}
Geiger and Rutherford had devised an electrical technique in order to count the individual $\alpha$-particles from radioactive sources, but Compton scattering was a far more complex process. Bothe and Geiger discussed the Bohr, Kramers, Slater theory, and quickly agreed that the question (``Is a light quantum and a recoil electron simultaneously emitted in the elementary process, or is there merely a statistical relation between the two?'') would have to be decided experimentally, before progress could be made: ``That such a decision was possible Geiger and I agreed immediately [\dots].''\cite{Bothe1954}

According to the Bohr, Kramers, Slater theory, a recoil electron would only occasionally be emitted when matter was traversed by X-rays. This prediction was in sharp contrast with the Compton theory, according to which a recoil electron appears every time a light quantum is scattered. A crucial test of these two predictions was possible using an apparatus devised by Bothe and Geiger to search for the simultaneous appearance of a scattered photon and the recoil electron. The main components were two of Geiger's needle counters, which were able to detect individual electrons. The two counters were separated by a very thin window of aluminum foil. A beam of carefully collimated X-rays was passed through one of the counters near the window and parallel to it. The first counter, called the {\it e counter}, could detect the recoil electron of a Compton-scattering process. A secondary photon emitted in the direction of the window could pass it, and give rise to another Compton effect in a thin platinum foil placed a short distance behind the window. The recoil electron of this secondary Compton scattering would be detected in the second counter. Because it detected the photon, this counter was called the {\it $h\nu$ counter}. The whole setup was placed in a glass sphere filled with hydrogen at atmospheric pressure. The two counters were connected to separate electrometers, and their deflections were recorded side by side on fast photographic film. Bothe and Geiger defined as a coincidence an event in which both counters showed a signal within a time interval of 1\,ms. They measured the deflection times on the developed film with an accuracy of 0.1\,ms. Thus, for any coincidence event the resolution was 1\,ms with the accuracy 0.1\,ms.\cite{Bothe1954b}

The question was whether a signal of the $h\nu$ counter occurred in coincidence with a signal of the $e$ counter more often than expected by pure chance.\cite{BGexperiment} If so, the two signals were related to the same process, a primary Compton scattering in the $e$ counter. Based on the number of coincidences between the signals due to recoil electrons and to the scattered X-rays, Bothe and Geiger estimated that according to the Bohr, Kramers, Slater theory the chance was only 1 in 400 000 that as many coincidences should have occurred. The result was therefore consistent with the predictions of the light-quantum theory. ``It is therefore to be assumed that the concept of the light quantum possesses a higher degree of reality than assumed in their [Bohr, Kramers, Slater] theory.''\cite{BotheGeiger1925} Soon afterward, Compton and Simon used a cloud chamber and found that energy and momentum are conserved in the scattering of X-rays on electrons.\cite{ComptonSimon1925}

Bothe commented in his Nobel Lecture of 1954: ``The final result we obtained was that systematic coincidences do indeed occur with the frequency that could be estimated from the experimental geometry and the response probabilities of the counters on the assumption that, in each elementary Compton process, a scattered quantum and a recoil electron are generated simultaneously.''\cite{Bothe1954}
The empirical refutation of the Bohr, Kramers, Slater theory by Bothe and Geiger established beyond doubt the strict validity of conservation principles in elementary processes, confirmed the physical reality of radiation quanta, and was a crucial experiment because, for the first time, two elementary particles, the electron and the photon\cite{Lewis1926} were simultaneously detected using a coincidence method.

In 1925, when Geiger took a teaching position as professor of physics at the University of Kiel, Bothe succeeded him as director of the laboratory of radioactivity at the Physikalisch-Technische Reichsanstalt and continued to work on experiments on the nature of light. He devised an experiment to display the quantum aspects of electromagnetic radiation by the coincidence method, testing the possibility of the purely statistical validity of the conservation theorems in another case. The idea underlying this new experiment was to ascertain 
whether there were time coincidences between the signals of two needle counters arranged in close opposite position ($180^{\circ}$). Both counters were able to detect fluorescent radiation stimulated by $\alpha$-particles impinging on a copper or iron foil interposed between them.\cite{Bothe1926} With such a setup Bothe would  observe no coincidences  if the scattered X-rays consist of light quanta. The result of this experiment was that no systematic coincidences occurred, 
at least not with the frequency predicted by the Bohr, Kramers, Slater theory. Bothe stressed in the abstract of Ref.~\onlinecite{Bothe1926} that ``From now on the processes can be described with `light quanta'.''
As he recalled in his Nobel Lecture,\cite{Bothe1954aa} during this time Bothe had the good fortune 
of being able to discuss the problem constantly with Einstein, who suggested some 
experiments to him, even if they yielded no decisively new result.

The wave-particle problem was made clear in a short time. The Davisson-Germer 
experiment, whose results appeared in the April 16, 1927 issue of Nature,\cite{DavissonGermer1927}
demonstrated the wave nature of the electron, confirming the earlier hypothesis of Louis de 
Broglie. Putting wave-particle duality on a firm experimental footing, in combination with 
Compton's experiment, represented a major step forward in the development of quantum 
mechanics.
At that time Compton scattering, bremsstrahlung, and photoelectric effect, 
even though considered well-established processes involving light quanta and electrons, were still lacking a fully satisfactory quantum treatment.

Compton was awarded the Nobel Prize in Physics in 1927 ``for his discovery of the effect named after him'' sharing it with Wilson who had devised a ``method of making the paths of electrically charged particles visible by condensation of vapor.''\cite{ComptonNobel} In presenting Compton's achievements, Karl Siegbahn remarked how the new wave mechanics ``lead as a logical consequence to the mathematical basis of Compton's theory,'' and concluded his speech emphasizing that the Compton effect, in gaining ``an acceptable connection with other observations in the sphere of radiation,'' proved to be of ``decisive influence upon the absorption of short-wave electromagnetic --- especially radioactive --- radiation and the newly discovered cosmic rays.''\cite{Siegbahn1927}

At that time the scientific community, which was well aware of the existence of an extraterrestrial radiation characterized by an incredible penetrating power (hence the name {\it penetrating radiation}), took for granted that this radiation was $\gamma$-rays of very high energy. Gamma rays were the most penetrating form of radioactive radiation known at the time, and hence the belief that cosmic rays were ``ultra $\gamma$-rays'' became widespread. 

To appreciate how this view was refuted, we need to go back to the beginning 
of the 20th century when it was discovered that charged electroscopes continued to lose charge no matter how well they were shielded or distanced from radioactive sources. In trying 
to explain this residual conductivity, physicists, despite their initial reluctance, were led to the assumption of extraterrestrial radiation falling upon the Earth.

\section{Speculating on the origin of cosmic radiation}

After some initial suggestions that there was a form of radiation of extraterrestrial origin, 
from 1909 onward the variation of the residual conductivity with altitude was investigated to 
determine whether the observed effect was associated with radioactive contamination in the 
air or in the environment. The hypothesis was that the effect could be explained in terms of $\gamma$-rays (high-energy electromagnetic radiation) emitted by 
radioactive substances present near the surface of the Earth's crust. To test this 
point, measurements were made by climbing towers or high mountains, but electroscopes were 
found to lose their electrical charge everywhere. It appeared to be impossible to 
eliminate the influence of these rays, no matter how thick the lead plates encasing the 
instrument. Recording devices were placed higher and higher to clarify the role of radiation 
coming from the Earth.
A series of balloon flights started by Albert Gockel and Victor Hess showed that the 
radiation first dropped to a minimum and then increased considerably with height. During the spring and summer of 1912 Hess made seven balloon ascents to heights up to 5300\,m and concluded that ``a radiation of very high penetration power enters our atmosphere from above.''\cite{Hess} In 1913--1914 Kolh\"orster ascended to an altitude of 9200\,m 
and confirmed Hess' results finding that the air ionization had increased up to ten times its value at sea 
level. These experiments led to the hypothesis
that part of the ionization must be due to 
radiation of extraterrestrial origin for which Hess coined the name
{\it H\"ohenstrahlung} (radiation from above). 

During 1910 and
1911 measurements of the intensity of the radiation were also independently
made with a novel method by Domenico Pacini at the Regio Ufficio
Centrale di Meteorologia e Geodinamica in Rome, Italy. Pacini enclosed his
electroscope in a copper box and immersed it in the Tyrrhenian Sea
near Livorno, and in the Bracciano Lake near Rome, measuring a
significant decrease of the radiation compared to the surface of the
water. He concluded that ``a sizable cause of ionization exists in
the atmosphere, originating from penetrating radiation, independent of
the direct action of radioactive substances in the soil.''\cite{Pacini1912}
Unfortunately Pacini died in 1934, and his work remained  generally unknown to
the international scientific community.\cite{PaciniDeangelis}

The conclusion of Hess and Kolh\"orster's experiments in the early 1910s 
was not immediately accepted and there was much controversy between 1914 
and 1926.\cite{Controversy} 
Millikan, who was director of the Norman Bridge Laboratory at the California Institute of 
Technology and had received the Nobel Prize in 1923 for his work on the elementary charge 
of electricity and on the photoelectric effect, was not convinced by prewar European results. 
Starting in the winter of 1921--1922 he decided to do his own experiments in 
collaboration with Ira Bowen.\cite{SelfRecording}
It initially appeared that their observations did not provide evidence for the existence 
of very penetrating rays of extraterrestrial origin, and Millikan rejected the idea of such a radiation. New absorption experiments done 
at Alpine Lakes at different altitudes during the summer of 1925 finally convinced Millikan 
and a most of the scientific community that the source of these rays was beyond the 
atmosphere. 
Millikan presented these results during a talk on November 9, 1925 at the 
National Academy of Sciences. The results showed ``that the rays do come in 
definitely from above, and that their origin is entirely outside the layer of atmosphere 
between the levels of the two lakes.''\cite{MillikanCosmicRays} At that time, Millikan coined the term ``cosmic rays'' for the mysterious radiation, a name which after over 80 years is still used, despite the evolution of the field.

An initial difficulty in accepting the existence of this new kind of penetrating radiation was 
probably related to the fact that the only known 
``high-energy'' processes were radioactive decays of nuclear nature, and the 
``radiation from above'' was found to have many times the penetrating power of the most 
penetrating radioactive radiations then known, the $\gamma$-rays from radium C.

In 1925 Millikan suggested that the penetrating rays, which he was 
convinced were high energy $\gamma$-rays entering the atmosphere isotropically from space, could be 
produced by the formation of helium from hydrogen, or by the capture of an electron by a 
positive nucleus. 
At the time only two processes were known during which radiation quanta lost energy 
interacting with matter: Compton scattering and ionization, in which a photon is absorbed with consequent emission 
of an electron from an atom.
Up to the beginning of the 1930s, electrons and ionized 
hydrogen were the only known elementary particles that were building blocks for atomic nuclei.
In the following years Millikan and G.\ Harvey Cameron studied the absorption of cosmic rays with 
the aim of determining the spectrum of cosmic ray energies. They found that their absorption curves in water were very similar to hard 
$\gamma$-ray absorption curves, a result also obtained by Erich Regener.\cite{GalisonRusso}

During the summer of 1927 Millikan obtained new results with the use of electroscopes 
``eight times more sensitive than those the authors had theretofore used.''\cite{MillikanCameron1928} Millikan tried to attribute his observations to the 
superposition of the absorption of a few identifiable ``ultra $\gamma$-rays.'' These results yielded an exponential law of absorption which corresponded roughly to 
the data when the observations were analyzed as a sum of contributions of 
exponential components corresponding to different $\gamma$-ray wavelengths and which, according 
to Millikan, showed that ``[\dots] the cosmic-ray spectrum consists of 
definite bands [\dots].'' The simplest hypothesis, to which he had already called attention 
in 1925, was that cosmic rays could be identified with the energy of formation 
of more complex nuclei from direct encounters of protons and high-speed electrons. His series of measurements 
extended up to a ``$\gamma$-ray'' energy of about 200\,MeV. Millikan presented his ``band theory'' 
on April 23, 1928 at the National Academy of Sciences.\cite{MillikanBands}
In speculating on the origin of the cosmic rays, Millikan and Cameron presented evidence ``to show that cosmic rays 
do not originate in the stars, but only in the depths of space.''\cite {MillikanCameronOrigin1928} In presenting a detailed analysis of these results  they challenged Eddington's and Jeans' ideas  about the possibility of an origin of the ultra $\gamma$-radiation in cosmic processes of matter formation in red giant stars.\cite{JeansEddington}

In other articles published during 1928\cite{MillikanCameron1928aa} Millikan continued to provide theoretical justifications for his theory, according to which cosmic rays were the ``birth cry'' 
of atoms in space being born in the form of $\gamma$-rays from the energy freed in the synthesis of 
heavier atoms through fusion of primeval hydrogen atoms. This idea had cosmological implications: ``The observed properties of cosmic rays, indicating 
that the creation of the common elements occurs only in interstellar or intergalactic space, 
suggest the possibility of avoiding the `W\"armetod' and of regarding the universe as already 
in `the steady state'.''\cite{MillikanCameronBands1928} Millikan's theory would soon have to tackle difficulties stemming from the quantum 
treatment of the absorption processes he had treated based on his ``fast and loose atomic 
physics.''\cite{Galison1987}

\section{Testing the nature of cosmic rays}

During the 1920s the main preoccupation of scientists working on cosmic rays was to establish their extraterrestrial origin. The variation of cosmic-ray intensity with altitude, their absorption and dissipation, and even their cosmic origin were of interest. Still there was the problem of the nature of cosmic rays, which did not attract general attention, probably because of the widespread belief that the astonishingly penetrating cosmic rays could not be anything else but $\gamma$-rays of very high energy. That view is reflected in Siegbahn's presentation speech.\cite{Siegbahn1927a}

While Millikan continued his nearly ``religious''  campaign to support his speculation 
on the origin of cosmic rays,\cite{Galison1987a} new instruments were being developed that would soon start 
a revolution in experimental cosmic ray research and transform it into a completely new field. A novel counting tube had just been developed in Kiel by Geiger and his 
pupil William M\"uller who announced their invention on July 7, 1928, during a meeting 
of the German Physical Society.
This device very soon appeared to respond to external 
radiation, which Geiger and M\"uller strongly suspected to be cosmic radiation.\cite{Trenn} At the time, they left open the question whether all the spontaneous counts should be attributed to cosmic rays.\cite{GeigerMuller1928} 

Electroscopes of the type used in earlier 
work on cosmic rays could detect only the combined ionizing effects of many
particles, as was true also of the ionization chambers used to investigate the absorption of cosmic radiation in lead and other materials. The ionization chambers consisted of gas-filled chambers containing 
two electrodes across which a potential was applied, high enough to remove the ions 
before recombination occurs. The resulting current was proportional to the number of ion pairs directly produced by all the charged particles traversing
the gas in a given time interval. As for the needle 
counter developed by Geiger,\cite{GeigerCounter1913} it was able to count single charged particles, but hardly suited to cosmic-ray studies. It counted over too 
small a volume, was not very stable, and was not accurate enough to study radiation whose intensity 
was as small as that of cosmic rays. The new Geiger-M\"uller counter had improved sensitivity and performance, and was capable of counting single ionizing particles, even if it could not tell anything about their identity or energy, except that they needed to have sufficient energy to penetrate the walls of the counter. It consisted of a closed vessel containing a gas and a pair of electrodes, but allowed for larger volumes and differed from the previous gas detectors in its parameter values (gas, type and pressure, and electric field). A charged particle
crossing the gas started a large ionization (discharge), and a current, which
in turn produced a voltage drop across a resistance in an external circuit.
The latter could be amplified to trigger a mechanical counter. Unlike the
ionization chamber, the relaxation time of the Geiger-M\"uller counter was 
short so that particles separated by short time intervals could be resolved.

Gamma-rays were known to ionize through the intermediary of secondary charged particles. Therefore, it was expected that cosmic $\gamma$-rays traveling through matter would 
be accompanied by secondary electrons resulting from the Compton effect, which were presumed to be the ionizing agent recorded by the measuring instruments. Conventional cosmic-ray absorption, 
as measured by placing an absorber above the detector, would reflect the $\gamma$-ray 
absorption and would not be affected by the properties of the secondary radiation. A direct 
study of this corpuscular radiation could thus clarify the nature of cosmic rays. 

At the end of July of 1928, an informal conference ``On $\gamma$- and $\beta$-ray 
problems,'' was held in Cambridge, UK. On July 12, Geiger had written
 to Rutherford: ``We have a counter now, which counts $\beta$-rays 
over an area of 100\,cm$^{2}$ and perhaps more. It is extremely sensitive [\ldots]. The counter 
must be screened from all sides with very thick iron plates to cut down the natural effect, 
which otherwise amounts to 100 throws [of the electrometer] per minute or more."\cite{Trenn1986}

During the Cambridge conference, the Russian physicist Dimitri Skobeltzyn working at the  Leningrad Physicotechnical Institute  presented some photographs of what he thought to be cosmic-ray tracks in his Wilson chamber. 
Following his remarks, Geiger announced that Bothe and Kolh\"orster ``were working on a 
method to register cosmic rays by the coincidence of pulses in two Geiger-M\"uller counters, and that 
they hoped to be able to study the penetrating power of the rays by this method."\cite{Skobeltzyn1985} 
By the time Geiger and M\"uller presented their final paper at the 90th meeting of German 
Scientists and Physicians in Hamburg in September 1928, they were reasonably sure that the 
spontaneous discharges were at least in large part  due to cosmic rays.\cite{Trenn1986}
In the meantime, Bothe's investigations
had stimulated Kolh\"orster, who was working in Geiger's laboratory at the 
Reichsanstalt as a permanent guest, to place two of the Geiger-M\"uller counters 
side by side into a beam of $\gamma$-rays. He argued that the recorded coincidences showed that ``only one and the same secondary 
electron had gone through both counters." Kolh\"orster, who had remarked that ``such a hard $\gamma$-radiation 
such as the H\"ohenstrahlung'' would generate secondary Compton electrons approximately in the 
same direction as the primary rays, succeeded in observing about three-fold more coincidences 
along the vertical with respect to the horizontal direction. Thus, it appeared that the coincidence method 
could be applied to detect the low-energy secondary electrons of the H\"ohenstrahlung. 
On  October 31, 1928, Kolh\"orster sent a note to Die Naturwissenschaften in which he wrote that the method was being applied to ``radiation from above."\cite{Kolhorster1928}
That was the first step into submitting the $\gamma$-ray hypothesis to a crucial
experimental test.

On the same page in the same issue of Die Naturwissenschaften, where
Kolh\"orster had published his short report on the first attempt 
to apply the coincidence method for the detection of the secondary electrons generated by the ``ultra $\gamma$-Strahlung,'' a note by Bothe and Kolh\"orster appeared.\cite{BotheKolhorster1928} They reported on their 
measurements of the absorption of those secondary electrons by recording the coincidences between two superimposed Geiger-M\"uller counters interleaved with lead plates of increasing thickness. From this arrangement they argued 
that coincidences could be produced only by individual ionizing particles crossing both counters. The experiment provided further evidence of 
the enormous potential of the coincidence method. Instead of establishing the 
simultaneous occurrence of two particles, as was done in the Bothe-Geiger experiments, 
it proved essential to follow the motion of one single particle amid many
simultaneous ionization effects by looking at a particular coincidence signal occurring in counters set according to a convenient geometry. As a result of their coincidence observations, Bothe and 
Kolh\"orster also reported the observation of ionizing particles penetrating 1\,cm of lead and 
concluded that their penetrating power could not be ascribed to the usual $\beta$-rays. 
The most natural conclusion was that these particles were secondary electrons 
generated by the ``ultra $\gamma$-radiation.'' 

In the meantime Skobeltzyn published a full report on his 
research, which was received on February 23, 1929 by Zeitschrift f\"ur Physik. 
Skobeltzyn used the just published Klein-Nishina theory of the Compton effect based on Dirac's 
relativistic equation for the electron,\cite{KleinNishina1929}
and concluded that the tracks of ionizing particles in his Wilson chamber were secondary electrons produced by 
the ``Hesschen ultra $\gamma$-Strahlen.''\cite{Skobeltzyn1929}

However, by April 1930 Bothe and Kolh\"orster took a very different position. In a two-page preliminary report in Die Naturwissenschaften, Bothe and Kolh\"orster, after presenting the main results of their research, reached the astonishing conclusion (which 
they emphasized): ``We think that the meaning of the whole result must be that the 
\textit{H\"ohenstrahlung} at least as far as the up to now observed evidence shows, is not a 
$\gamma$-radiation, but a corpuscular radiation.''\cite{BotheKolhorster1929}

On  June 28, 1929, a final paper was submitted. In the beginning of the article Bothe and Kolh\"orster clearly highlighted the 
problem of cosmic rays:
\begin{quote}
\small
``Research into the high-altitude radiation has so far taken a strange course, for the 
most diverse features of the radiation, such as intensity, distribution, absorption and scattering, and even its origin, are investigated and debated, whilst the really essential question regarding the nature of the high-altitude radiation has hitherto found 
no experimental answer. The main reason for this is that owing to the 
low intensity and great penetrating power of these rays it is not possible directly to examine a screened beam of rays. If the view that the high-altitude radiation is a very 
hard $\gamma$-radiation has until now been universally preferred, this has only been because 
of the enormous penetrating power, which would be more difficult to explain by a corpuscular 
radiation [\ldots]. The essential problem is thus the following: Is this corpuscular radiation 
to be understood as secondary to a $\gamma$-radiation, as has been customary, or does it 
represent in itself the high-altitude radiation? To answer this question it is especially 
important to know the penetrating power of the corpuscular radiation. If this corresponds 
to that of the high-altitude radiation itself, it will strongly support the hypothesis that 
the high-altitude radiation itself has a corpuscular nature. If on the other hand the 
corpuscular radiation is markedly softer than the high-altitude radiation, the latter 
must be a $\gamma$-radiation which produces the corpuscular radiation by impact on matter. 
We have now performed an experiment to determine the penetrating power of the supposed 
secondary electrons, proceeding from the above $\gamma$-ray hypothesis. It turned out that 
this experiment made it possible to decide between the above-mentioned interpretations."\cite{BotheKolhorsterArticle1929} 
\normalsize
\end{quote}

According to the proponents of the primary $\gamma$-radiation hypothesis, the Compton recoil 
electrons of very high-energy photons would have more than enough energy to traverse both 
counters' walls. However, they 
should be completely absorbed by a very thin absorber between the counters. 
When Bothe and Kolh\"orster recorded coincidences with and without a 4.1\,cm thick gold 
block between the counters (see Fig.~1), the results were startling. To their surprise, 
the gold layer produced only a moderate decrease in the counting rate, which meant that 
76\% of the charged particles present in the cosmic radiation near sea level could 
penetrate 4.1\,cm of gold. The penetrating power of the ionizing particles, the 
``fast electrons'' responsible for the double coincidences, appeared to be almost as high as the ``ultra radiation'' itself.\cite{DoubleCompton} 

\begin{figure}[h!]
\begin{center}
\includegraphics[width=11 cm]{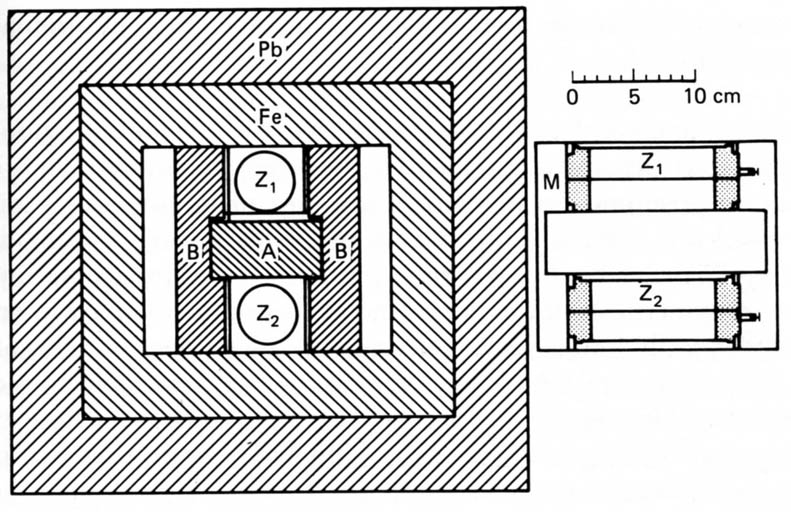}
\caption{The experiment of Bothe and Kolh\"orster in Ref.~\onlinecite{BotheKolhorsterArticle1929}. Coincidences between counters $Z_{1}$ and 
$Z_{2}$ are produced by cosmic-ray particles traversing both counters. Signals could be produced only by cosmic radiation because the shielding prevented signals due to radioactivity. Observations were 
made by recording the simultaneous pulses with and without a 4.1\,cm thick gold absorber, 
chosen because of its high density. Coincidences were recorded by connecting the wires of the 
two counters to two separate fiber electrometers, which were imaged over a moving photographic 
film. Time resolution was about 1/100\,s.}
\end{center}
\end{figure}

Regarding these experiments, Bothe and Kolh\"orster could not help say that the ``ultra radiation'' was not of a wave nature, but consisted of corpuscles which they thought to be high velocity electrons.\cite{electrons} They concluded that the observed ionizing particles were not of secondary nature, but were the 
quanta of the local cosmic radiation itself. They thus suggested that the primary cosmic 
radiation consisted of ionizing particles, and the ionizing particles observed near sea 
level were among the primary particles that were capable of traversing the 
atmosphere. 

Their counter experiments did not prove that the primary 
rays were corpuscular, and their latter hypothesis turned out to be incorrect, 
but at the time it had a role in defining the astrophysical and the physical aspects of cosmic-rays. It was still difficult to accept that corpuscles could have the high energies that were necessary to enable them to penetrate the atmosphere. There was still another objection: If the primary rays were charged particles, the Earth's magnetic field should have an influence on their distribution, but in the late 1920s convincing proof of the 
expected intensity dependence of cosmic radiation on the magnetic latitude had not yet been found. 

\section{A new technological window: the Rossi coincidence circuit}

Bothe and Kolh\"orster's research on local cosmic radiation and their hypothesis 
of its corpuscular nature appeared in November 1929.\cite{BotheKolhorsterArticle1929} Their landmark paper aroused the curiosity of the young Italian physicist 
Bruno Rossi: ``[it] came like a flash of light revealing the existence of an unsuspected world, full of mysteries, which no one had yet begun to explore. It soon became my overwhelming ambition to participate in the exploration."\cite{Rossi1966}

After studying at the University of Padua, Rossi received a \textit{summa cum laude} 
degree in physics at the University of Bologna at the end of 1927, and in early 1928 he 
started his academic career at the University of Florence. This university had been founded in 1924, only a few years before his arrival. 
The Physics Institute was located on the Arcetri Hill, about 3 kilometers out of town, at 
least 150\,m above the level of the city, and close to Villa Il Gioiello, the residence of Galileo 
Galilei during the last period of his life. It was directed by Antonio Garbasso, who had been 
trained at the German school of Heinrich Hertz in Bonn and of Herman von Helmholtz in 
Berlin. The 23-year-old Rossi was eager to work on an experimental project addressing 
``the discovery of some secret of nature,'' and ``the fundamental
problems of contemporary physics.''\cite{Rossi1990} Rossi knew Millikan's theory, which ``was certainly a 
fascinating hypothesis, for it suggested an answer to two of the most fundamental problems 
of contemporary science: that is, the origin of the atoms of different elements and the origin 
of cosmic rays.'' His doctoral student, Giuseppe Occhialini, ``was fascinated by it,'' and called his attention to a paper 
where Millikan  had presented his ideas, but Rossi was skeptical: ``[\dots] in my opinion, 
the interpretation of the experimental observations on which Millikan had built his theory 
was not convincing. On the other hand, I did not see how, being an experimentalist, I 
could contribute significantly to the current speculations about the origin of cosmic 
rays."\cite{Rossi1990a}
A research program such as Millikan's would require a lot of 
money, and a similar project was not within the young Italian
physicist's reach. Moreover, the subject had been explored mostly by senior scientists 
over a period of years. Rossi was critical of Millikan's interpretation of cosmic rays as high-energy $\gamma$-radiation generated in the depths of space,  but 
at the same time he admitted having ``accepted uncritically'' this idea.\cite{Rossi1981}

The $\gamma$-ray assumption appeared to be disproved by Bothe and Kolh\"orster's 
experiment,\cite{BotheKolhorsterArticle1929} even if it would be misleading to claim that it had showed the corpuscular nature 
of cosmic rays. Bothe and Kolh\"orster were aware of the ``potential pitfalls'' 
involved in their conclusions about the nature of the primary cosmic-rays. According to Rossi, 
``There was in fact one serious objection to the conclusions reached by Bothe and 
Kolh\"orster. The interpretation of their experiments was based on an arbitrary extrapolation 
of the known properties of photons and electrons at low energies. It was conceivable, for 
example, that the energies of cosmic-ray photons might be much greater than those calculated 
from their mean free path according to the equation of Klein and Nishina, which was known to 
be valid for energies of the order of 1\,MeV. If such were the case, the secondary electrons 
would have had a greater range, more of them would have penetrated the gold block between the 
counters, and they might have produced much the same coincidence effects as a primary 
corpuscular radiation."\cite{Rossi1966a}

Bothe 
and Kolh\"orster's experiment\cite{BotheKolhorster1929} represented the first attempt to determine the 
nature of cosmic rays, and contributed to focusing physicists' interest on local 
radiation, that is, the radiation found at the place where measurements were made. This kind of 
physics could be performed in a terrestrial laboratory. Rossi quickly realized how the Geiger-M\"uller 
countercould become the key to open ``a window upon a new, unknown territory, with unlimited 
opportunities for explorations."\cite{Rossi1981a}
The opening of ``new windows on the universe'' would be a \textit{leitmotif} in 
Rossi's scientific life.

Research on cosmic rays had joined with investigations concerning 
electronic counting devices. As Rita Brunetti, one of Rossi's teachers in Bologna, 
observed, ``the history of physics instruments is exactly fitting in with the history of physics."\cite{Brunetti1936}
The advent of the Geiger-M\"uller counter marked the end of the first period of 
cosmic-ray research, which now, ``became truly a branch of physics."\cite{Rossi1982}
A new era was beginning during which the Geiger-M\"uller counter was for a 
long time the keystone of cosmic-ray physics. Rossi's own life as a scientist was intertwined with all its remarkable developments and applications. 

Excited and full of enthusiasm for the possibilities raised by the Bothe and Kolh\"orster,\cite{BotheKolhorster1929} Rossi immediately set to work with the help of his students, 
particularly Giuseppe Occhialini and Daria Bocciarelli, and his colleague Gilberto 
Bernardini. Rossi knew that he had very limited means at his disposal, but the 
novelty of the research topic, and the low cost of the research tools were the key ingredients  
of his excitement. He was 24 years old when ``one of the most exhilarating periods'' of his life 
began.\cite{Rossi1990b} 
In just a few months he built his own Geiger-M\"uller counters, devised a new 
method for recording coincidences, and began some experiments. According to Occhialini, the Geiger-M\"uller counter ``was like the Colt pistol in the American frontier: the `great leveler,' 
namely an inexpensive detector, within the reach of every small laboratory, requiring only good 
scientific intuition and experimental skill to get useful results in the new fields of nuclear 
physics and cosmic rays."\cite{Occhialini}
And the Geiger-M\"uller technique, which Rossi introduced to Italian physics, 
played a crucial role in the discovery of the radioactivity induced by neutrons and 
for the related discovery of nuclear reactions brought about by slow neutrons made by Enrico 
Fermi and his group in Rome during 1934.\cite{Leone2005}
In retrospect, the rapidity with which Rossi launched into experimental physics and performed remarkable research in a new field, is remarkable. As a beginner, Rossi had to compete with wily old foxes who knew their 
job well, both as experimenters with a solid background in physics. Also, they had worked for a long time in research institutions characterized by an 
outstanding theoretical and experimental tradition such as those in Germany. But within a few weeks the 
first counter was in operation, and Rossi could now tackle the coincidence technique, 
which was at the core of the Bothe-Kolh\"orster experiment, and with incredible insight and skill 
succeeded in fully developing the capabilities of the method. 

On November 12, 1929 Bothe submitted a paper which appeared in the January 1930 issue of 
Zeitschrift f\"ur Physik, describing a method for registering simultaneous pulses of 
two Geiger counters.\cite{Bothe1930}
The brief electric current produced by each counter when detecting a particle was sent to one of the grids of a special 
two-grid valve, the other grid being connected to the second counter. Bothe regulated the device so 
that the tetrode, a Telefunken RES 044, was triggered only when receiving a signal from 
both counters. He obtained an electric signal that he could register, visualize, and 
listen to, thanks to a loudspeaker.
After Lee De Forest invented the triode in 1906, valves found applications outside 
of communication technologies when Jacob Kunz in 1917 outlined a method by which a 
photoelectric current could be amplified using a triode, thus making a photoelectric 
cell more useful as a photometer.\cite{Kunz1917}
But scientific use of valves as amplifiers of weak signals was not considered 
until 1924, when the Swiss physicist Heinrich Greinacher had the idea of combining vacuum tube 
amplifiers with needle counters and later with a small 
ionization chamber with the aim of amplifying the signal.\cite{Greinacher} These were probably among the 
earliest applications of vacuum tubes for purposes other than handling electromagnetic 
oscillations.

In the same period, Rossi's improved version of the coincidence circuit, submitted on  February 7,
1930, appeared in the April issue of Nature (see Fig.~2).\cite{Rossi1930} He had the simple but 
ingenious idea of using the triode as an automatic switch: when the grid has a positive 
potential, current is flowing, the ``switch'' is closed; when a negative signal is sent on 
the grid, current is not flowing, and the ``switch'' is open. In his mixed 
arrangement,  consisting of Geiger-M\"uller counters and valves, a pulse would be sent by 
the output terminal of the circuit only when signals were received at a specified number of input 
terminals within an assigned time interval. 

\begin{figure}
\begin{center}
\includegraphics[width=14 cm]{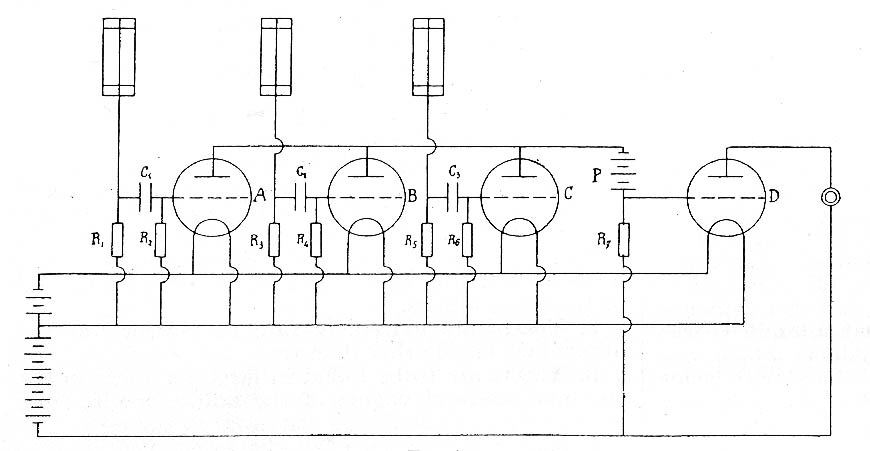}
\caption{Rossi's coincidence circuit appearing in  Ref.~\onlinecite{Rossi1930}. 
The selecting resistance on the right  ($R_{7}$) is adjusted in relation to the plate resistance of 
a single vacuum tube in its normal state so that when negative pulses are impressed upon the grids 
of {\it all but one} of the tubes, the current carried by the other tubes is still sufficient to 
produce an $iR_{7}$ drop across the resistance nearly equal to the full battery potential. If, on 
the other hand, negative pulses are received on the grids of {\it all} of the tubes, the $iR_{7}$ 
drop across the resistance disappears and the grid of the output tube receives a positive pulse, 
which results in the recording of a count.}
\end{center}
\end{figure} 

In showing the electric diagram for three counters whose positive electrodes were one-to-one coupled to the grid of three valves, Rossi noted that ``in normal conditions these grids have a zero potential; whenever a discharge 
occurs they become negative, thus interrupting the current flow.'' The anodes of the valves were 
kept at a near zero potential by a very large resistance, while a fourth valve had a slight 
negative potential thanks to an auxiliary battery. This 
potential 
varies very little when only one or two counter tubes are working, while ``it undergoes 
a sudden rise when, for the simultaneous working of the three counter tubes, the current is 
interrupted in all the three valves.'' The consequent variation of the anode current was acoustically 
detected by a telephone.
Rossi also remarked that the arrangement of his circuit was symmetrical in regard to the 
inputs to counters, a condition which was ``not fulfilled in the circuit of Prof.\ Bothe, because 
the grids of the two-grid valve have rather different characteristics.''\cite{Rossi1930}

The time resolution of Rossi's threefold coincidence circuit offered a tenfold improvement and was conceptually different from Bothe's method, which employed a single tetrode vacuum tube and could register only twofold coincidences. The presence of three counters or more in coincidence in the $n$-fold version of the circuit, greatly reduced the rate of chance coincidences, thus allowing observations with increased statistical weight. Moreover, the time correlation among associated particles crossing different counters could be established.

If the Geiger-M\"uller counter was an instrument of precision, being a tool more discriminating than the 
ionization chamber, particularly regarding directional effects, the electronic coincidence circuit 
radically changed our view of cosmic rays. The possibility of arranging more counters, in whatever geometrical
configuration, opened new possibilities of investigation. In particular, it would 
soon prove fundamental for studying secondary effects such as the production of new particles 
from interactions between cosmic rays and matter. These arrangements, were the precursors of the 
AND logic circuit later used in electronic computers. 

The coincidence method became of vital importance for all experiments with several electronic detectors, and Rossi's circuit was soon widely adopted around the world. The coincidence
method, with specific electronic coincidence circuits, also became a powerful tool in nuclear physic research, namely for nuclear spectroscopy, a new field pioneered by Bothe and his collaborators.\cite{BotheNuclear} The Institut f\"ur Physik at the Kaiser Wilhelm-Institut f\"ur medizinische Forschung in Heidelberg, under Bothe's direction since 1934, became an important international center for nuclear research. Yet, Bothe kept an interest in the secondary effects of cosmic radiation, which remained a part of the rich program of his Institute.\cite{BotheHilgert} In 1954 Bothe was awarded the Nobel prize
for physics (shared with Max Born) ``for the coincidence method and his
discoveries made therewith.''\cite{BotheNobel} 

\section{Probing cosmic rays via geomagnetic effects}

While the exploration of the nucleus was entering a more mature era, both experimentally and theoretically, cosmic rays were becoming an object of research. In particular, problems relating to the interaction of cosmic rays with matter, which belonged more properly to the field of ``radiation and nuclear physics'' could throw light on the properties of nuclei of different elements.

There still remained the problem of understanding the nature and charge of the cosmic-ray particles. With his innovative electronic setup Rossi was paving the way for future 
experimental practice in the field. This innovation turned into 
the first part of his early research program aimed at demonstrating the corpuscular nature and properties of cosmic 
rays, in contrast to the theory that considered them as high frequency $\gamma$-rays.

In particular, Rossi stressed that precious information of on the charge and the velocity of cosmic ray particles could be extracted by magnetic deviation phenomena. Besides analyzing the corpuscles' reaction in the presence of an electromagnet to verify that they carry an electric charge and to determine its sign (Fig.~3),\cite{RossiMagnetic} Rossi 
used the Earth's magnetic field as part of a natural spectrometer. If primary cosmic rays were charged 
particles, they would also be affected by the geomagnetic field before entering into the terrestrial 
atmosphere. 

\begin{figure}
\begin{center}
\includegraphics[width=5 cm]{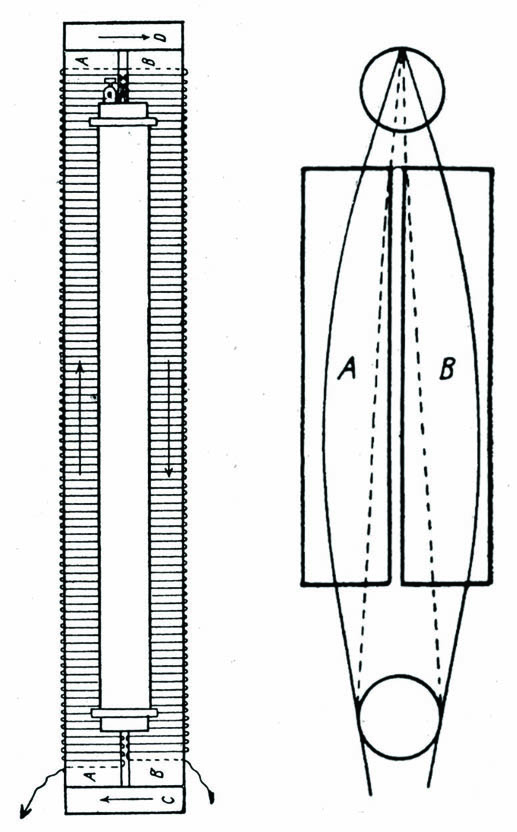}
 \caption{View of the device suggested by Luigi Puccianti and used by  Rossi for the experiment of magnetic deviation.\cite{RossiMagnetic} The core 
of the magnet consisted of two iron plates \textit{A} and \textit{B},  the two armatures \textit{C} 
and \textit{D} closed the magnetic circuit. The wire carrying the magnetizing current was wound 
round the plates in a single layer;  the closed induction lines pass through the core, as indicated by the arrows, clockwise or 
counter-clockwise, according to the direction of the magnetizing current. Above and below the magnet 
there are two tube counters; their axes are horizontal and parallel to the direction of magnetization. Particles passing through the upper counter are concentrated toward the lower counter or deflected away from it depending on the sign of their charge.  Rossi measured the coincidence rates with opposite magnetic fields, and found at first a very little difference in favor of positive particles, an effect much smaller than the expected one. In a second experiment the difference was within the limits of the experimental errors. On the right, the side view shows the trajectory of particles with a ``wanted'' charge sign (solid lines), and of ``unwanted'' ones (dashed lines). On the left, the top view shows the direction of the magnetic field and the position of the  top G-M counter.}
\end{center}
\end{figure}

A latitude effect was actually expected indicating a lower intensity of cosmic rays near the equator, where the horizontal component of the geomagnetic field is stronger. A slight variation of the intensity  had been observed by Jacob Clay in 1927 and 1929, resulting from experiments made carrying ionization detectors onboard ships that traversed an extensive latitude range stretching from Netherland to Java.\cite{ClayLatitude} The observed effect could possibly be attributed to the different meteorological conditions present in such different locations.  Negative  results had been found by Bothe and Kolh\"orster,\cite{BotheKolhorsterLatitude}  and by Millikan\cite{MillikanLatitude} during recent experiences carried out respectively between Hamburg and Spitzbergen and between Bolivia and Canada.

On July 3, 1930 Rossi submitted a paper in which he conjectured the existence of a second geomagnetic effect
 which would be revealed by an asymmetry of the cosmic-ray intensity with respect to the plane of the geomagnetic meridian, with more particles coming from east or west, 
depending on whether the particle charge was negative or positive.\cite{Rossi1930EW} 
He tried to detect the  east-west effect  at Florence using two counters arranged in a vertical plane as a ``cosmic-ray telescope'' to select particles incident from a restricted solid angle. However, within experimental error, the results were negative.  

During one of his frequent visits to Rome, Rossi had a discussion with Enrico Fermi concerning the influence of the Earth's magnetic field on the distribution of cosmic-rays intensity. Fermi pointed out that according to Liouville's theorem the intensity should be the same in all allowed directions. The intensity distribution of the incident particles near the Earth could thus be determined if it was known which directions of incidence are ``forbidden'' by the magnetic field. On the basis of this result, they re-examined the data from the Florence experiment, as well as the results of the experiments of Clay and others on the latitude effect,\cite{ClayLatitude,BotheKolhorsterLatitude,MillikanLatitude} and concluded that an abnormally large atmospheric absorption provided the most likely interpretation of all the available data. Assuming the value of the atmospheric losses indicated by their analysis, they concluded that in the vicinity of the equator the east-west effect should have been clearly observable. Within a few hours after Rossi's arrival at the Physics Institute of via Panisperna, Fermi and Rossi completed a manuscript, the only one they wrote together, even though they continued their dialogue on cosmic rays until Fermi's premature death in 1954.\cite{RossiFermi}

On the strength of this prediction Rossi decided to speed up the organization of the already planned expedition to Eritrea, in the proximity of Asmara, 
the capital of the Italian colony, at a geomagnetic latitude of 11$^{\circ}$ 30' N, where the asymmetry was expected to be larger, and at an altitude up to 2370\,m, sufficient to observe cosmic rays of comparatively small energy. But funds did not arrive soon enough, and in 1933, when he was about to 
leave for Eritrea,
experiments proving the existence of a difference in the intensity of cosmic rays between east and 
west were announced in two letters sent to Physical Review by Luis Alvarez 
and Compton,\cite{AlvarezCompton} as well as by Thomas Johnson.\cite{Johnson} Soon after, the effect was also observed by Rossi 
and Sergio De Benedetti in Eritrea.\cite{EastWestEffect} 

Rossi and De Benedetti reported another result of their observations. They found that ``[\dots] once in a while the recording equipment is struck by very extensive showers of particles, which cause coincidences between counters, even placed at large distances from one another."\cite{showers} This  was the first  explicit description of extensive air showers, which a 
few years later were ``rediscovered'' and studied in detail by Pierre Auger and his collaborators,\cite{Auger} and which would become a principal theme of Rossi's postwar cosmic-ray research at the Massachusetts Institute of Technology.

The study of the asymmetry predicted by Rossi since the summer of 1930 confirmed the corpuscular 
hypothesis of the nature of the primary radiation and revealed that the primaries, important 
in producing effects near sea level, were mostly positively charged. At the time, the latter 
was a most unexpected result ``because those of us who had supported the corpuscular theory 
were convinced --- more, I must admit, from prejudice than because of a logical argument --- that 
the primary particles would turn out to be electrons, and would therefore be negatively rather 
than positively charged."\cite{Rossi1990d}

At the same time, stimulated by discussions with Rossi, Compton planned a detailed world-wide survey, which divided the Earth into nine zones and teams, with all investigators using identical large ionization chambers. The cosmic-ray intensity was measured in 69 stations distributed over 
a wide range of geomagnetic latitudes and longitudes and at many elevations above sea level. This world-wide project involving more than 60 physicists greatly extended Clay's 
earlier observations done between Holland and Java. It established beyond doubt the reality of the latitude effect and proved that at least a significant fraction of the primary cosmic rays are charged particles and are subject to the influence of the Earth's magnetic field. The existence of the two  geomagnetic effects now ``called as insistently for the recognition of charged particles in the primary radiation as its supposed absence had formerly necessitated a denial of the contribution by such particles."\cite{Swann} In conjunction with the coincidence experiments, these observations definitely contradicted Millikan's theory that all the charged particles 
detected at sea level were secondary generated within the Earth's atmosphere by cosmic 
$\gamma$-radiation.\cite{RussoDeMaria1989} 

\section{Unveiling the existence of the soft and the hard components of cosmic rays}

In parallel with this line of research, Rossi had also explored the interaction between particles 
and matter. In trying to obtain direct information on the mysterious radiation, Rossi set his ``traps'' by wisely arranging metal screens and circuits of counters 
according to different geometrical configurations.

Though unaware of it, Rossi was acting as a ``particle hunter.'' He used procedures that future particle physicists employed, such as 
selecting a beam of particles, having it collide with a target, and observing what happens when it passes through an alternate sequence of counters and screens (detectors). During 1930--1932 Rossi discovered that cosmic-ray particles could pass through enormously thick matter, including up to a meter 
of lead (Fig.~4).\cite{Rossi1mblei} His findings gave evidence of the astonishing energies associated with cosmic rays. 

	\begin{figure}[htbp]
\begin{center}
\includegraphics[width=9 cm]{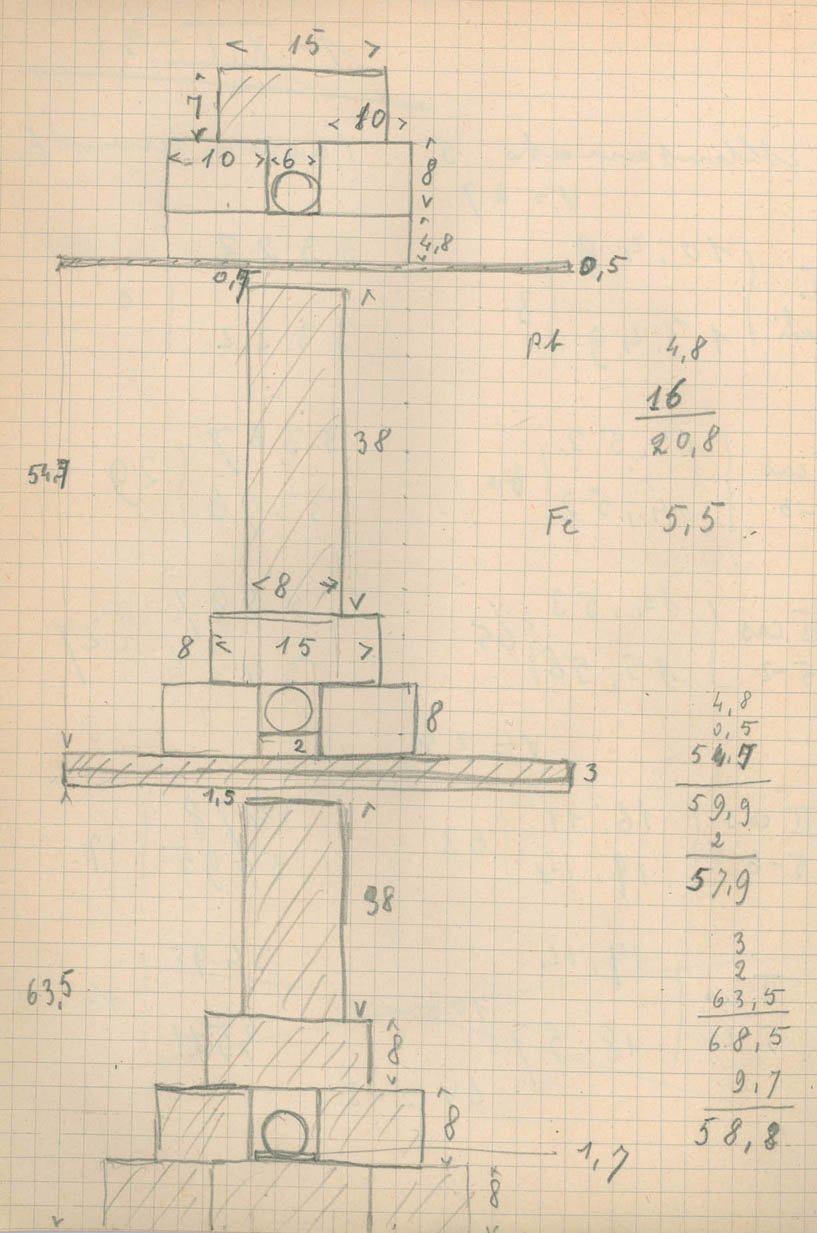}
\caption{Original drawing of Rossi's experimental arrangement which proved that the cosmic rays  contained particles capable of traversing 1 m of lead.\cite{Rossi1mblei} Coincidences were recorded by the three counters $C_{1}$, $C_{2}$ and $C_{3}$ separated by a thick lead layer. Threefold 
coincidences instead of twofold coincidences were used, in order to reduce the chance 
coincidence rate. B. Rossi, Notebook,  Institute Archives \& Special Collections, Massachusetts 
Institute of Technology Libraries, Cambridge, Massachusetts.}
\end{center}
\end{figure}

Using arrangements of counters such as the one in Fig.~5 (a), in conjunction with his threefold coincidence
circuit, Rossi also discovered that individual cosmic rays traversing a metal absorber, which was placed above the counter array, 
often generated secondary particles. The coincidence rates reported in the first experiment appeared so contrary to common sense to the editors of Die Naturwissenschaften to whom Rossi had submitted a short note, that they refused to publish it. The paper was later accepted by Physikalische Zeitschrift,\cite{Rossi1932}  after Heisenberg vouched for its credibility.

\begin{figure}[h!]
\begin{center}
\includegraphics[width=16 cm]{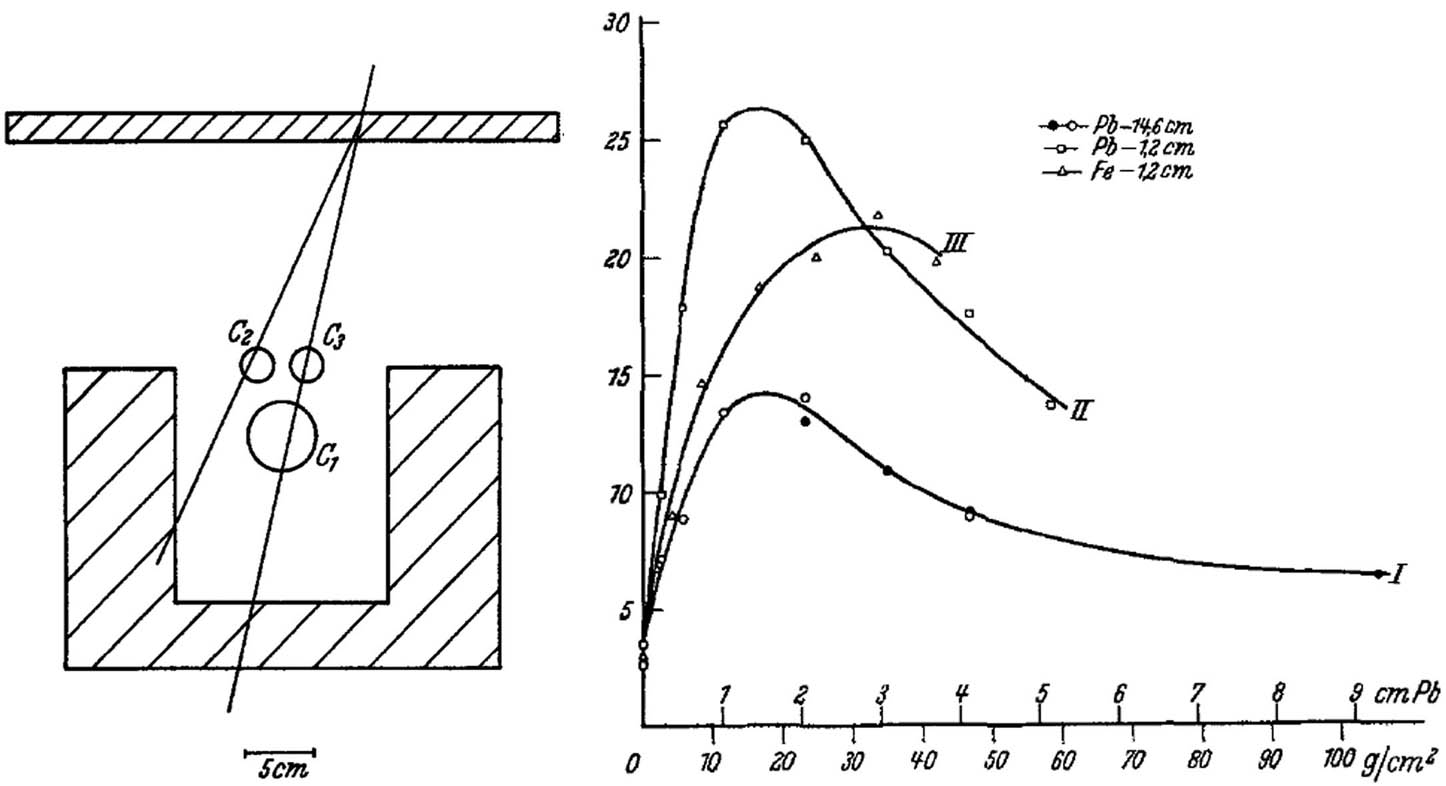}
\caption{Hard and soft components of cosmic radiation at sea level.\cite{Rossi1933} (a) A typical Rossi's setup is shown. It consists of a triangular array of Geiger-M\"uller counters enclosed in a lead box. The top wall could be removed and replaced by a screen of different material, thickness, and distance from the counter array. (b) The coincidence rate is displayed as a function of the thickness of the lead or iron screen placed above the counters. Curves I and II refer to measurements taken with lead screens at distances of 14.6 and 1.2\,cm, respectively, above the counters. Curve III refers to measurements taken with an iron screen at a distance of 1.2\,cm above the counters. Each curve results from the superposition of two terms, with the soft component mainly contributing to the coincidence rate at small thickness of the screen, and quickly dropping off (rapid rise and initial decrease in the shower production) while the hard component contribution continues up to a large thickness, and decreases very slowly (long tail of the lead curve). The soft component contribution increases with the screen atomic number, as seen by comparing curves II and III obtained with lead and iron screens, respectively, at a fixed distance of 1.2\,cm from the counter array.}
\end{center}
\end{figure}

Rossi's investigation pointed to the existence of two components in cosmic rays
at sea level: a ``hard" component, able to pass through 1\,m of lead
after being filtered through a 10\,cm thick lead screen; and a ``soft" 
component, generated in the atmosphere by primary cosmic rays and able to
generate groups of particles in a metal screen before being stopped. Rossi's experiments showed that hard and soft rays 
were fundamentally different in character and did not differ merely in their energy. The results were summarized in curves similar to those displayed in 
Fig.~5(b), since then known as {\it Rossi transition
curves}, which represented the coincidence rate of coherent groups
of particles versus the thickness of the absorber placed at a given
distance above the counter array.\cite{Rossi1933} 
These results could be interpreted, for the example of a lead absorber, by assuming that a soft secondary component of cosmic rays was at the origin of a ``tertiary radiation producing most of the threefold coincidences observed with 1\,cm lead.''\cite{Rossi1932a} 

The possibility of the existence of a secondary soft radiation in the atmosphere had been suggested 
by Rossi in 1931.\cite{Rossi1931} He made comparative measurements on the number of coincidences  inclining more and more the plane containing the axes of two counters when a lead screen 9 cm thick was introduced between them. By increasing the filtering effect of the atmosphere he expected "to find a progressive hardening of the corpuscular radiation.Ó However, the experimental results did not confirm his expectations. With the two counters
placed vertically the one above the other, he measured a lower absorption
than the one found when the plane of his Geiger-M\"uller telescope was oriented at an angle
of 60$^{\circ}$ to the vertical line. Rossi concluded, therefore, that
``the slant rays are softer and not harder than the vertical ones. This
result may be accounted for by assuming that the corpuscular rays generate
in the atmosphere a soft secondary corpuscular radiation, and that the
relative amount of the latter is larger in an inclined direction than in
the vertical one.''

At this time the Geiger counter coincidence 
measurements, and thus the method Rossi and his collaborators adopted in their 
investigations, combined with the particular arrangement of counters and metal shields, were especially criticized by Millikan: 
``I have been pointing out for two years in Pasadena seminars, in the Rome Congress on 
nuclear physics in October, 1931, in New Orleans last Christmas at the AAAS meeting, 
and in the report for the Paris Electrical Congress, that these counter 
experiments never in my judgment actually measure the absorption coefficients of 
anything. I shall presently show that no appreciable number of these observed ionizing 
particles ever go through more than 30\,cm or at most 60\,cm of lead and yet both 
Regener and Cameron and I have proved that the cosmic rays penetrate through the 
equivalent of more than 20 feet of lead. These figures cannot both be correct without carrying with them the conclusion that the primary rays at sea level and below 
are not charged particles."\cite{Millikan1933} 
The following year, during the London Conference on Nuclear Physics in October 1934, 
Millikan, Bowen, and Neher, and also Anderson and Neddermeyer, still postulated that 
the primary radiation consisted mostly of photons. Millikan had not even mentioned Rossi's experiments, which were the main 
target of his criticism, as Rossi, since the beginning of his activity in cosmic-ray 
research, had been one of the principal proponents of the corpuscular hypothesis, 
and had dared to criticize his ``birth cry theory'' for the origin of cosmic rays 
during the first international conference on nuclear physics organized in Rome by Fermi in 1931.
\cite{RossiMillikan}

\section{Showers of particles}

In the summer of 1932 groups of several particles were found in a large fraction of the 
photographs taken in Manchester by Patrick Blackett and Rossi's former pupil, 
Giuseppe Occhialini, using the coincident discharge of a ``Rossi counter telescope'' to trigger 
a cloud chamber inside a magnetic field.\cite{LeoneRobotti} Instead of using the usual method of random expansion of the chamber,  they placed 
Geiger-M\"uller counters above and below a vertical cloud chamber, so that any ray passing 
through the two counters would also pass through the chamber, triggering its expansion. 

They waited for cosmic rays to arrive and ``take their 
own cloud photographs.'' Some pictures revealed an ``astonishing variety and complexity'' of multiple tracks, which seemed to have a common origin above the chamber, and which Blackett and Occhialini called ``showers.'' Because of 
the appearance of groups of particles diverging downward, they suggested 
that ``those bent to the left are negatively charged and those bent to the 
right are positively charged."\cite{BlackettOcchialini}

``It is interesting to note," as Blackett recalled later, ``that the development of the 
counter-controlled cloud chamber method, not only attained the original objective of 
achieving much economy in both time and film, but proved to have the unexpected 
advantage of greatly enhancing the number of associated rays photographed. This was so 
because the greater the number of rays in a shower of cosmic ray particles, the greater 
the chance that the counter system controlling the chamber would be set off. As a result 
the larger showers appeared in the photographs far more frequently relative to 
single rays than they actually occur in nature. This property of bias towards complex 
and so interesting phenomena has proved one of the most important advantages of the 
counter-controlled method."\cite{Blackett1948} 

In their second paper, Blackett and Occhialini pointed out that the occurrence of these tracks was a well known feature of cosmic radiation and was ``clearly related to the various secondary processes occurring when penetrating radiation passes through matter.'' They also credited Rossi for having been the first to investigate these secondary particles using counters.\cite{BlackettOcchialini2} 
The effect revealed by Rossi's experiments with counters (see Fig.~5) was now also definitely established visually, and it was confirmed that the top metal screen in Rossi's setup acted as an absorber of cosmic radiation as well as a source of particle showers.\cite{Bhabha1933}

In the meantime, Carl D.\ Anderson, one of Millikan's young collaborators, sent off a short manuscript which appeared in the September 1932 issue of Science. For some time he had measured the energies of the charged particles produced by cosmic rays by measuring the track curvature in a cloud chamber, roughly of the same size as that used by Blackett and Occhialini,  and randomly expanded in a magnetic field up to 2.1\,T. 
After discussing different interpretations of the observed tracks, he concluded that ``For the interpretation of these effects 
it seems necessary to call upon a positively charged particle having a mass 
comparable with that of an electron."\cite{Anderson1932}
As later recalled by Anderson, ``It was not immediately obvious to me, however, as to just what the detailed mechanism was."\cite{AndersonPositron}

After a detailed examination and 
interpretation of their photographs, Blackett and Occhialini were able to 
confirm the interpretation proposed by Anderson. They also placed the positive 
electron into theoretical perspective, arguing that because there are no 
free electrons in light nuclei, the negative and positive electrons in the 
showers must have been created during the process. Thus, based on the 
clear evidence of the process known as 
``pair-creation" provided by their counter-controlled chamber (see Fig.~6),
Blackett and Occhialini confirmed Dirac's relativistic
theory of the electron.\cite{BlackettOcchialini3} They were the first 
to expound the pair formation mechanism derived from experiments, which was the 
process underlying the formation of electromagnetic showers, one of the most striking 
facts of the phenomenology related to cosmic rays. 

\begin{figure}[htbp]
\begin{center}
\includegraphics[width=10 cm]{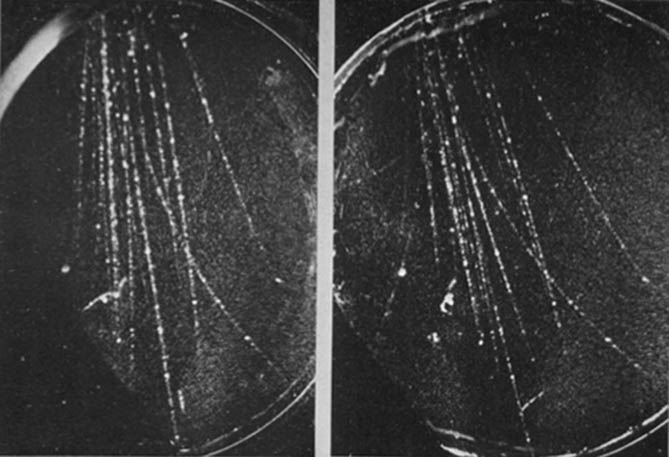}
\caption{Shower of about 16 separate tracks showing formation of  electron-positron couples produced by Blackett and  Occhialini's counter-controlled cloud chamber. The chamber of diameter 13 cm and depth 3 cm was arranged with its plane vertical and two G-M counters, each 10 cm by 2 cm, were placed one above and one below the chamber so that any ray which passed straight through both counters had also to pass through the illuminated part of the chamber.  The counters were connected to a valve circuit arranged to record only simultaneous discharges of the two counters. The whole chamber was placed in a water cooled solenoid containing a field of 0.3 T. From P. Blackett and G. Occhialini, Ref.~\onlinecite{BlackettOcchialini}, plate 22, facing p. 722. The method relied on the very assumption that the penetrating radiation at sea level was electrically charged, and tracks were found on 80\% of the photographs.}
\end{center}
\end{figure}

The study of showers was greatly helped by the theory of the electromagnetic 
cascade appearing in early 1937 in simultaneous papers by J.\ Franklin Carlson and J.\ Robert 
Oppenheimer,\cite{CarlsonOppenheimer}  and by Homi Bhabha and Walter Heitler,\cite{BhabhaHeitler} which provided a natural and simple explanation on the basis of the quantum electrodynamics cross-sections calculated by Bethe and Heitler.
Electrons were known to ionize and to lose energy by emitting photons. 
Photons were now known to form electron-positron pairs 
and to give rise to recoil electrons by the Compton effect. Thus, a shower 
could arise from a cascade process where an electron radiates a high-energy 
photon, and the photon forms an electron pair or produces a 
Compton electron. The new electron, as well as the original one, would radiate more photons and so on, until all of the primary energy was dissipated 
in ionization. For the first time it was demonstrated that radiation could
transform into matter, and that the energy required to produce such a pair 
was $2mc^{2}\approx1$\,MeV, where $m$ is the electron mass. 
A good experimental test of the theory was required, based on a large amount of data 
resulting from the observation of many interactions. By comparing Rossi's curves with the ones expected from their theory, 
Bhabha and Heitler found quantitative agreement between theory and experiments, at least with respect to the soft component of cosmic rays.\cite{BhabhaHeitler} 

More detailed tests followed, confirming the success of the theory. Already in the mid 1930s, an extension of the coincidence method 
was suggested by the necessity to select events in which some counters were 
discharged while others were not. The main reason for directly measuring anti-coincidence rates, (for example coincidences between counters $BCD$ not accompanied by a discharge in  $A$) rather than the difference between two coincidence rates ($BCD$ and $BCDA$) measured separately, is the large difference between the statistical errors of the experimental results in the two cases. The ``anti-coincidence technique,'' which 
Rossi used in an elegant experiment in 1938--1939 with Lajos J\'anossy to 
investigate the photon component of the local cosmic radiation,\cite{RossiJanossy} was soon widely used, and became an established practice 
for all sorts of later experiments (Fig.~7).

\begin{figure}[htbp]
\begin{center}
\includegraphics[width=14 cm]{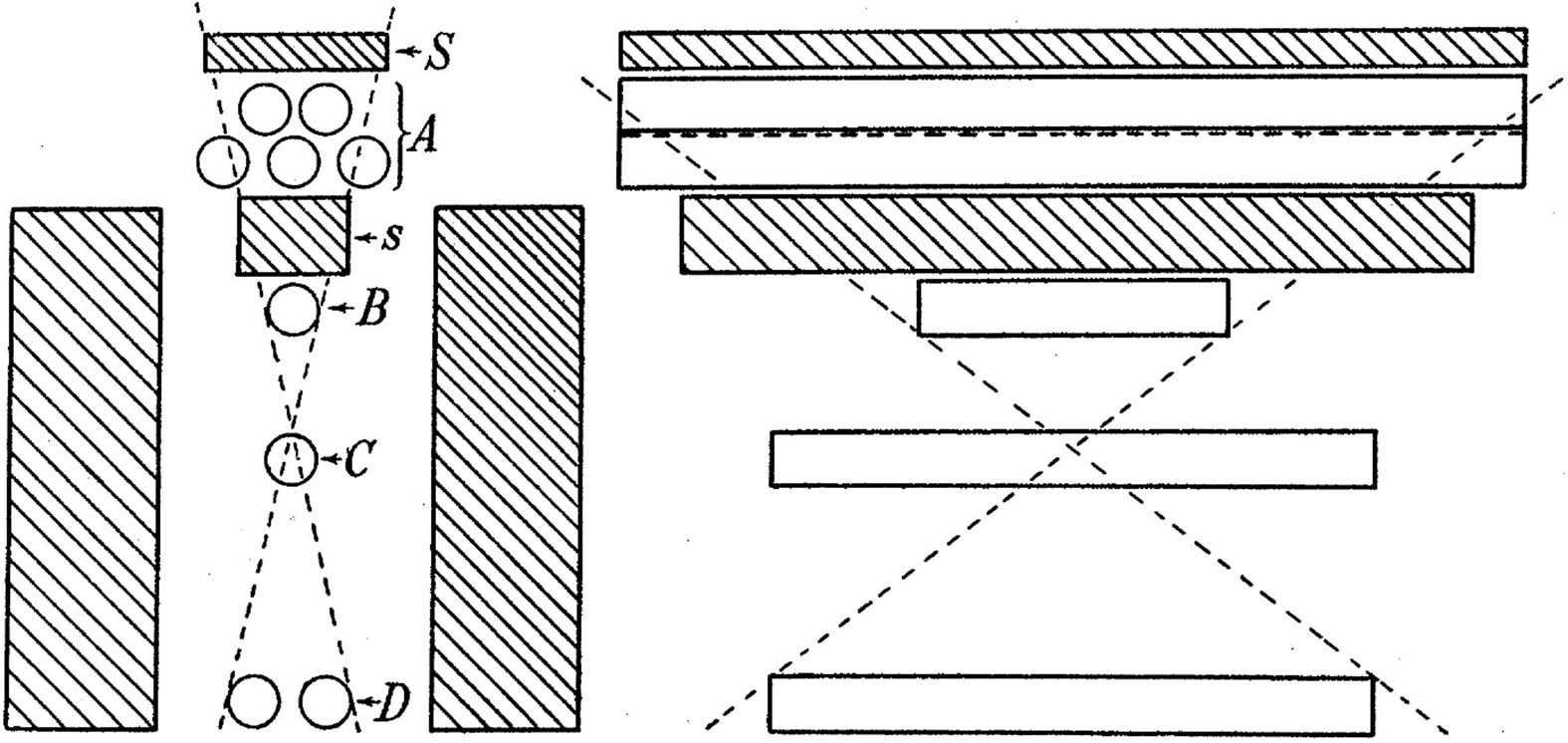}
\caption{An early application of the anti-coincidence method devised by Rossi in an experiment carried out with J\'anossy studying the photon component of cosmic radiation.\cite{RossiJanossy} The principle was to detect the electrons generated by photons (Compton effect or pair production) in a metal plate.  The counter battery {\it A} (60 cm long, 3 cm diameter) covers the whole solid angle subtended by the counters {\it B} (20 cm long, 3 cm diameter), {\it C} and {\it D} (40 cm long, 3 cm diameter). Each single ionizing particle passing through {\it B}, {\it C} and {\it D} must have passed through {\it A} as well, unless produced by some non-ionizing agent between {\it A} and {\it B}. Incident photons  traversing the absorber {\it S} (60$\times$7 cm$^{2}$) will thus go through counters {\it A} without discharging them because they do not ionize the gas. Some of them will then generate pairs of secondary electrons which will discharge counters {\it B}, {\it C} and {\it D}. Thus the arrival of a photon will be identified by a coincidence between counters {\it B}, {\it C}, and {\it D} {\it not} accompanied by a discharge in counters  {\it A (anticoincidence}  ${BCD-A}$).}
\end{center}
\end{figure}

By 1941, a review of what was known theoretically about cosmic ray behavior was prepared by Rossi in collaboration with his student Kenneth Greisen.\cite{RossiGreisen}  The article focused on the interaction of cosmic rays with matter, the problem Rossi had faced since the beginning of his researches. This article came to be known as ``The Bible'' in the cosmic-ray community, and was used for many years as a reference text by particle physicists.

In that same period Pierre Auger and Roland Maze discovered that cosmic-ray particles separated even by distances as large as 300\,m arrived in time coincidence. The phenomenon of extensive air showers showed that there existed particles with an energy of $10^{15}$\,eV at a time when the largest energies involved in natural radioactivity processes were just a few MeV.\cite{Auger1} The conclusions reached by Auger and his colleagues depended on the statistical analysis of many events, aided by the newly developed theory of electron-photon cascades.

\section{Cosmic rays and the birth of particle physics}

The electromagnetic cascade theory left completely open the problem of the hard component of cosmic rays, which had been revealed 
by Rossi's 1932 experiments showing that 
a large proportion of the corpuscular radiation found at sea level 
still had the capability of traversing more than 1\,m of lead (see Fig.~4). 
Carlson and Oppenheimer's conclusion was ``[\dots] 
either that the theoretical estimates of the probability of these processes 
are inapplicable in the domain of cosmic-ray energies, or that the actual 
penetration of these rays has to be ascribed to the presence of a penetrating 
component other than electrons and photons.'' According to them, 
if these were not electrons, they must be ``particles not previously known 
to physics."\cite{CarlsonOppenheimer1}

The enigma was soon solved by 
Anderson and Seth Neddermeyer. In an article received on 30 March 1937 they suggested that ``there exist particles 
of unit charge but with a mass (which may not have a unique value) larger 
than that of a normal free electron and much smaller than that of a proton."\cite{AndersonNeddermeyer1937} Soon after, Jabez C.\ Street and Edward C.\ Stevenson, who used a three-counter telescope to select the penetrating particles and a lead filter for removing shower particles, published the first photograph of a mesotron:\cite{StreetStevenson} ``[\dots] for many people it was the most impressive piece of evidence for the new particle."\cite{GalisonMuon} From measurement of its ionization and momentum Street and Stevenson deduced a value of about 130 electron masses, compared to the currently accepted value 
of $105.658367\pm0.000004)$\,MeV/$c^2$, that is about 200 times the mass of an electron.\cite{pdg2010}

The new particle was called
by various names, including barytron, yukon, mesotron,
and meson, the last two names being suggested by its intermediate 
mass value. The name yukon came from identifying 
it as the particle predicted by Hideki Yukawa in 1935,\cite{Yukawa} to serve as the mediator 
of the strong nuclear force binding together the nucleus, in analogy to the 
exchange of the massless photon in quantum electrodynamics. 

From 
analogy with $\beta$-decay it was assumed that the mesotron would be unstable and disintegrate spontaneously into an electron and a
neutrino.\cite{Bhabha1938} For the first time physicists were dealing with the spontaneous instability of an elementary particle. Establishing the reality of such a process --- and the accurate determination of the mesotron's mean lifetime --- became one of the outstanding problems of cosmic ray research. Immediately after his arrival in the U.S.,\cite{RossiEmigration} Rossi became involved as well.\cite{RossiMesotronReview} Between 1939 and 1941, the problem was solved with a series of experiments which are exemplars of Rossi's style.

To test the hypothesis of the spontaneous decay of mesotrons, Rossi and his collaborators, Norman Hilberry and Barton Hoag, compared the absorption of the mesotron component of cosmic rays in air and in carbon. The mesotrons were detected by the coincidences of three Geiger-M\"uller counters arranged in a vertical plane at
four different heights from Chicago up to the top of Mt.\ Evans, Colorado (4300\,m). The ionization losses 
in the carbon absorber were equivalent to those in the atmosphere above. 
The quantity that was experimentally determined was the average range
before decay $L$, where $dz/L$ represents the fractional number of 
mesotrons decaying while traveling the distance $dz$, and $L=p\tau_0/m$
following the relativistic transformation for time intervals, where
$p$ and $m$ are the momentum and the rest mass of the mesotron. With 
$m$ = 80\,MeV/c$^2$ and $<1/pc>_{\rm av}=0.8\times10^{-3}$\,MeV$^{-1}$, the 
lifetime turned out to be $\tau_{0}=2\times10^{-6}$\,s. The attenuation of
the hard component was greater in the atmosphere than in a carbon absorber
with equivalent ionization losses. The difference was interpreted as due to the spontaneous decay of the mesotrons in flight.\cite{RossiMesotron1} These 
measurements provided the first unambiguous demonstration of the anomalous 
absorption of mesotrons in the atmosphere.

To 
eliminate all sources of doubt, Rossi, together with David B.\ Hall, designed a new experiment to test the dependence of the disintegration probability on momentum, 
thus providing another check of the decay hypothesis, and also to verify 
the relativistic transformation for time intervals. Mesotrons with a 
momentum between 440 and 580\,MeV/c or more were selected by recording 
only mesotrons that traverse a given thickness of lead and are stopped by 
a certain additional absorber. An anti-coincidence circuit greatly helped with
regard to the statistical precision. Measurements were taken at Denver 
(1616\,m) and Echo Lake (3240\,m) resulting in a fairly monokinetic group 
of mesotrons with a lifetime $\tau_{0}=2.4\times10^{-6}$\,s.\cite{RossiMesotron2} 

Given different results obtained by other experimenters it 
was desirable to compare the decay of mesotrons within two narrow 
momentum intervals, (452---586\,MeV/c) and (903---1044\,MeV/c). 
The measurements were done under similar experimental conditions as in
the previous year, resulting in 
$\tau_{0} = (2.8\pm 0.3)\times10^{-6}$\,s and $\tau_{0} = (2.9\pm0.7)\times10^{-6}$\,s,
respectively. By combining the new data with the re-analyzed values of Rossi and 
Hall, a value $\tau_{0}=(2.8\pm0.2)\times10^{-6}$\,s was obtained.\cite{RossiMesotron3}

After the last mountain expedition, Rossi believed that no significant additional progress could be made by further measurements of the same kind. Moreover, they provided only an order of magnitude for the mean time before decay of mesotrons at rest: ``The measurement of the decay of mesotrons in flight had given us the ratio $\tau_{0}/m$ of the mesotron mean life to its mass, but the mass of mesotrons, at that time was not known with any accuracy. Moreover, despite all precautions, we could not discount the possibility of systematic errors in our measurements. Thus, although the order of magnitude of the mean life was by then well established, the uncertainty in its exact numerical value was probably greater than the 7\% statistical error of our final result."\cite{Rossi1983}

At that time, the only observation of mesotrons decaying at the end of their range was found in several cloud-chamber tracks photographed in 1940 by E.\ J.\ Williams and G.\ E.\ Roberts.\cite{Williams} The first successful direct measurement of the mean lifetime of the decay process was performed by Franco Rasetti, who had been a leading member of Fermi's group in Rome and who had moved to Laval University in Canada in 1939. The absorption of a mesotron by a block of aluminum or iron was recorded by a system of coincidence and anti-coincidence counters. Another system of counters and circuits registered the delayed emission of a particle, which was interpreted as the disintegration electron associated with the absorbed mesotron. The apparatus enabled him to determine the time distribution of the emitted particles and hence the mean life of the decay process, which was found to be ($1.5\pm0.3)\,\mu$s,\cite{Rasetti1941} in agreement with the value deduced from the atmospheric absorption of fast mesotrons at different altitudes previously found by Rossi and his collaborators.\cite{RossiMesotron2}

Rossi took up the challenge, and in the last approach to the decay problem he tackled the necessity of measuring time intervals between the discharges of Geiger-M\"uller counters ranging from a fraction of one to several $\mu$s and invented a new electronic circuit, the first of the time-measuring devices, later known as ``time-to-amplitude'' converters. With his collaborator Norris Nereson, Rossi used this ``electronic chronometer'' to increase the selectivity and the statistical accuracy of the method considerably by recording all decay electrons and measuring the time interval between the arrival of each mesotron and the emission of the electron arising from its decay.\cite{RossiMesotron4} Observation of several hundred decays made it possible to plot experimental decay curves, such as that showed in Fig.~8, which was the first measured decay curve of an elementary particle.

\begin{figure}[htbp]
\begin{center}
\includegraphics[width=15 cm]{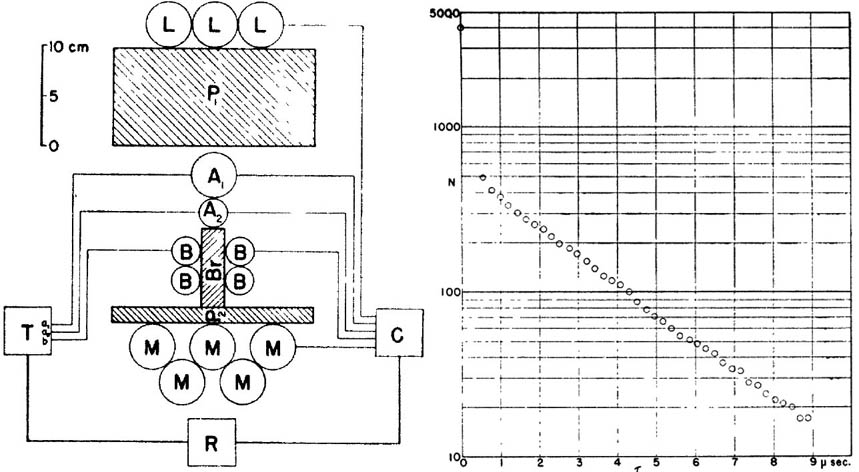}
\caption{(Left) Rossi and Nereson's experiment for measuring  the decay curve of mesotrons at rest.\cite{RossiMesotron4}  An electronic circuit  {\it C} records the anticoincidences produced  by mesotrons which, after traversing  \textit{L}, \textit{A}$_{1}$, \textit{A}$_{2}$, and the screen $P_{1}$ come to rest  in the brass plate \textit{Br} and subsequently decay, producing an electron which discharges one of counters B, thus failing to discharge counters {\it M}  (anticoincidence  $LA_{1}A_{2} B-M$).  Such event is detected by the circuit $C$, whose resolving time is long compared with the lifetime of mesotrons. The heart of the experiment is the electronic `clock' {\it T} which measures the time intervals between the arrival of
mesotrons on the absorber \textit{Br} and the emission of their decay electrons.
Apart from the time lag of the counters, this time is equal  to the life of 
the mesotrons in the brass plate \textit{Br}, when the anticoincidence is caused by a 
disintegration process. (Right) Experimental disintegration curve of mesotrons. The abscissa $\tau$ is the delay recorded by the time circuit, the ordinate is the logarithm of the number \textit{N} of anticoincidences accompanied by delays larger than the corresponding abscissa. The experimental points lie on a straight line as closely as one can expect considering the statistical fluctuations. Hence the disintegration of mesotrons follows an exponential law as does any disintegration process.}
\end{center}
\end{figure}

The value of $2.3\pm0.2)\,\mu$s was determined in extraordinary agreement 
with the currently accepted value of $(2.197034\pm 0.000021) \,\mu$s.\cite{pdg2010} Rossi and Nereson used such a value together with the value of $\tau_{0}/m$ determined with Hall and obtained $ m =160$ electron masses, which was within the limits of 160 and 240 according to the most reliable measurements of the time obtained with different methods.

By 1942 the evidence for the decay of the mesotron at the end of their range had changed from the first two cloud-chamber tracks photographed by Williams and Roberts to the curve presented by Rossi and Nereson, which contained thousands of decay events and showed an exponential decay with a lifetime of about 2 $\mu$s.

The style and elegance of these experiments, which were the 
first demonstration of the time dilation of moving clocks predicted by 
Einstein's theory of relativity, had their counterpart in the investigations on the fate 
of mesotrons coming to rest in matter, which during the war years were carried out in 
Rome by Marcello Conversi, Ettore Pancini, and Oreste Piccioni. The Rome experiment was initially inspired by Rasetti's approach using the delayed 
coincidence method to signal the decays of stopped particles, with one
important variant, the introduction of the magnetic-lenses array already 
used by Rossi in 1931 (see Fig.~3 and Ref.~\onlinecite{RossiMagnetic}), which focused particles 
of one sign of charge and defocused those of the opposite sign.\cite{CPP} 

What was immediately seen to be a crucial experiment, was later described by Luis Alvarez during his Nobel Lecture: ``[\dots] modern particle physics 
started in the last days of World War II, when a group of young Italians, Conversi, 
Pancini and Piccioni, who were hiding from the German occupying forces, initiated a 
remarkable experiment [\dots].''\cite{Alvarez1968}

The Italian physicists' data showed that the mesotron interacted with nuclei $10^{12}$ times more weakly than expected and could not possibly fulfill the role of the Yukawa particle, thus providing the first hints 
of a much more complex underlying reality. 

The riddle was definitively solved in 1947, 
when Cecil Powell, Cesare Lattes, and Occhialini at the University of Bristol 
identified a new particle in photographic emulsions exposed at high altitude.\cite{LPO} It became clear that the just discovered $\pi$-meson was the real Yukawa meson, 
and that the mesotron --- later dubbed the muon --- was the product of its decay. The muon was later classified as a lepton, like the electron. Both these particles do not feel the effects of the strong nuclear force. 

The field of high-energy particle physics started in the early 1930s by cosmic-ray investigations and continued into the 1950s, when accelerators became the dominant source of particles. By that time it had become clear that primary cosmic-ray particles colliding with nuclei of atoms in the atmosphere produce a cascade of secondary processes which during the early 1950s provided a whole ``zoo'' of new elementary and {\it strange} particles, leading to the rise of theories of the nature of matter. However, the way toward a satisfactory theory of fundamental particles and their interactions would prove long and tortuous.

The problem concerning the nature of the primary cosmic radiation was solved only in the early 1940s, when balloon experiments by M.\ Schein and co-workers carried out with complex Geiger-M\"uller counter arrangements showed that much of the primary particles consist of protons.\cite{Schein} These results were corroborated at the end of the 1940s by measurements taken at various altitudes and aboard B-29 airplanes using nuclear emulsions. It was thus found that cosmic rays contain nuclei of various elements including iron, whose energies amounted to many GeV. Protons make up about 90\% of the primary cosmic rays while helium nuclei ($\alpha$-particles) are nearly 10\%, and all other nuclei to around 1\%.\cite{Bradt} 

\section{Some concluding remarks}

At the end of 1920s particle physics was on the verge of emerging ``out of the turbulent confluence of 
three initially distinct bodies of research: nuclear physics, cosmic-ray studies, 
and quantum field theory,'' as Brown and Hoddeson remarked in their introduction 
to the proceedings of a symposium dedicated to the birth of particle physics.\cite{BrownHoddeson} At that time, Bothe's as well as Rossi's style of work was
highlighted as exemplary in establishing the ``logic'' research tradition,\cite{Galison1987a} which paved the way for future investigations. Starting from the discovery of the positron, the visual impact of cloud chamber images and later of nuclear emulsions, played a crucial part in the demonstration of the existence of the mesotron (the muon) and the $\pi$-meson. Up to the early fifties the ``image'' tradition continued to provide a series of ``golden events'' which became instrumental in establishing the existence of a rich subatomic world. 

At the same time, the logic tradition was instrumental in dealing 
with large samples of events connected to a single kind of phenomenon.   In devising more and more refined configurations made of counters and 
electronics components the followers of the logic tradition were creating a rich  experimental base, a connective texture  contributing to the  gradual understanding of long-unsolved mysteries 
regarding the nature and behavior of cosmic radiation. 

After the discovery of the positron, theoreticians began to pay much attention to cosmic rays as important to the development of theory, and cosmic-ray experimenters recognized quantum mechanics as an important tool, closely related to their experiments. The investigation of cosmic rays was facilitated by the fast coincidence method of multiple counter discharges developed by Rossi and by its application to the counter-controlled expansion of a cloud chamber by Blackett and Occhialini,  a perfect fusion of  ``image'' and ``logic''. These two methods were particularly useful for investigating the interaction of charged particles with matter, because an impinging charged particle is selected by the discharges of aligned counters in coincidence.
In the period when understanding was far from clear and quantum electrodynamics appeared to break down at the high energies involved in cosmic rays, experiments using counters became one of the fundamental tools for testing the new physics. The theory explaining shower formation and the identification of the hard component with a brand-new particle, the mesotron, eventually resurrected relativistic quantum field theory toward the end of the 1930s.\cite{BrownRechenberg}

During the 1940s, up to the advent of high-energy accelerators at the beginning of the 1950s, 
cosmic rays continued to play a leading role as a source of high-energy particles. In particular, they provided the great amount of data needed to test relativistic field theories, which came from the rich phenomenology of high-energy interactions involving particle creation and annihilation in these processes. The two instrumental traditions emerging from the early studies on the nature of cosmic rays continued to develop, each with its own practitioners and with different styles, sometimes competing, but often collaborating in devising new methods of particle detection that could provide evidence for the diversity of the subatomic world. 

Yet that was also the era of ``little science,'' the 
same science characterizing the simple tabletop experiments performed by Thomson and Rutherford at the dawn of the 20th century. As Rossi noted later ``[\dots] results bearing on fundamental problems of elementary particle physics could be 
achieved by experiments of an almost childish simplicity, costing a few thousand dollars, 
requiring only the help of one or two graduate students.''\cite{Rossi1983a}  A most remarkable change in the second half of twentieth century science was its growth in scale, scope, and cost. No wonder that in the 1980s, in an era of big science, and as a mature representative of the old generation, Rossi felt ``a lingering 
nostalgia'' for what he called the ``age of innocence of the physics of elementary 
particles.''

\begin{acknowledgments}
It is a pleasure to express my gratitude to Paolo Lipari for most precious discussions, to 
Carlo Bernardini for having the patience to 
read a preliminary version of this work, and to Alessandro De Angelis for some important remarks. I owe a special debt of gratitude to Maria Fidecaro for her generous support and for her most insightful comments during the final revision. I am also indebted to two anonymous referees for helpful criticisms and suggestions, and to Susan Leech O'Neal for her thoughtful linguistic revision of my paper.
\end{acknowledgments}


\begin{thebibliography}{5}
\bibitem{BGexperiment}Well aware that some coincidences could result from uncorrelated signals (chance coincidences), Bothe and Geiger also devised a simple method to evaluate their number. The evaluation was based on the number of events in which the two signals were at a time distance larger than 1\,ms, for example, between 5 and 6\,ms.
\bibitem{electrons}This conclusion is not surprising, considering that in the 1920s electrons and ionized hydrogen were the only known elementary particles to serve as building blocks for atoms.
\end{thebibliography}
\end{document}